\documentclass[twocolumn,aps,prl,longbibliography,superscriptaddress,amsmath,amssymb,superscriptaddress]{revtex4-2}
\usepackage[english]{babel}
\usepackage[utf8]{inputenc}
\usepackage{amsthm}
\usepackage{mathtools}
\usepackage{siunitx} 
\usepackage{braket}
\usepackage{xcolor}
\usepackage{graphicx}
\usepackage{soul}
\usepackage{adjustbox}
\usepackage{placeins}
\usepackage[T1]{fontenc}
\usepackage{lipsum}
\usepackage{csquotes}\usepackage[pdftex, pdftitle={Article}, pdfauthor={Author}]{hyperref} 

\begin{document}
\title{MoSi Superconducting Nanowire Single-Photon Detectors on GaAs for On-Chip Integration}
\author{M.~Erbe}
 \email{marcel.erbe@unibas.ch}
\affiliation{ 
Department of Physics, University of Basel, Klingelbergstrasse 82, CH-4056 Basel, Switzerland}%
\author{R.~Berrazouane}
\affiliation{
ID Quantique SA, Rue Eug\`{e}ne-Marziano 25, CH-1227 Gen\`{e}ve, Switzerland}%
\author{S.~Geyer}
\affiliation{ 
Department of Physics, University of Basel, Klingelbergstrasse 82, CH-4056 Basel, Switzerland}
\author{L.~Stasi}
\affiliation{
Group of Applied Physics, University of Geneva, CH-1211 Gen\`{e}ve, Switzerland}
\author{F.~van~der~Brugge}
\affiliation{
ID Quantique SA, Rue Eug\`{e}ne-Marziano 25, CH-1227 Gen\`{e}ve, Switzerland}%
\author{G.~Gras}
\affiliation{
ID Quantique SA, Rue Eug\`{e}ne-Marziano 25, CH-1227 Gen\`{e}ve, Switzerland}%
\author{M.~Schmidt}
\affiliation{ 
Lehrstuhl für Angewandte Festkörperphysik, Ruhr-Universität Bochum, DE-44780 Bochum, Germany}%
\author{A.~D.~Wieck}
\affiliation{ 
Lehrstuhl für Angewandte Festkörperphysik, Ruhr-Universität Bochum, DE-44780 Bochum, Germany}%
\author{A.~Ludwig}
\affiliation{ 
Lehrstuhl für Angewandte Festkörperphysik, Ruhr-Universität Bochum, DE-44780 Bochum, Germany}%
\author{F.~Bussi\`{e}res}
\affiliation{
ID Quantique SA, Rue Eug\`{e}ne-Marziano 25, CH-1227 Gen\`{e}ve, Switzerland}%
\author{R.~J.~Warburton}
\affiliation{ 
Department of Physics, University of Basel, Klingelbergstrasse 82, CH-4056 Basel, Switzerland}%

\date{\today}
            
\begin{abstract}
We report on MoSi-based superconducting nanowire single-photon detectors on a gallium arsenide substrate. MoSi deposited on a passivated GaAs surface has the same critical temperature as MoSi deposited on silicon. The critical temperature decreases slightly on depositing MoSi directly on the native oxide of GaAs. Hence, MoSi works well as a thin-film superconductor on GaAs. We propose that the amorphous structure of MoSi ensures compatibility with the GaAs matrix. Superconducting nanowire single-photon detectors (SNSPDs) are fabricated with MoSi on GaAs using a meander-wire design. The SNSPD metrics are very similar to those of devices fabricated with the same procedure on a silicon substrate. We observe a plateau in the response-versus-bias curve signalling a saturated internal quantum efficiency. The plateau remains even at an elevated temperature, 2.2~K, at a wavelength of 980~nm. We achieve a timing jitter of 50~ps and a recovery time of 29~ns. These results point to the promise of integrating MoSi SNSPDs with GaAs photonic circuits.
\end{abstract}

\maketitle

\section{Introduction}
Quantum photonics, the manipulation and detection of photonic qubits, is pursued increasingly on-chip. Typically, the waveguides consist of silicon or silicon nitride. GaAs is also a possibility \cite{Wang2014fq}. Individual InAs quantum dots create single photons in the GaAs waveguides with near-unity efficiency \cite{arcari_sollner_javadi_lindskov_hansen_mahmoodian_liu_thyrrestrup_lee_song_stobbe_et_al._2014,ba_hoang_beetz_midolo_skacel_lermer_kamp_hofling_balet_chauvin_fiore_et_al._2012}, and the photons can be highly indistinguishable \cite{Uppu2020wv}. Furthermore, GaAs is electro-optically active allowing for the creation of voltage-tunable switches and phase shifters \cite{Midolo2017ev}. Quantum photonic integrated circuits on this platform can be fabricated in a single lithographic step without the need for a hard etching mask \cite{midolo_pregnolato_kirsanske_stobbe_2015}. 

Quantum photonics requires not only single-photon sources and a waveguide circuit but also single-photon detectors with high efficiency, low dark count-rate, high repetition rate, and low timing jitter. Such metrics can be achieved presently only with superconducting nanowire single-photon detectors (SNSPDs) \cite{hadfield_2009}. These detectors are based on a superconductor switching from the superconducting to the resistive state by absorption of a photon \cite{Goltsman2001jh,natarajan_tanner_hadfield_2012,esmaeil_zadeh_chang_los_gyger_elshaari_steinhauer_dorenbos_zwiller_2021}.

It is advantageous to integrate the detectors with the on-chip waveguides. On the one hand, evanescent coupling of an in-waveguide photon to an on-waveguide detector can be highly efficient \cite{pernice_schuck_minaeva_li_goltsman_sergienko_tang_2012}. On the other hand, by removing the need for out-coupling to an external detector, photon loss can be much reduced. The challenge lies in the heterogeneous nature of the materials used for the waveguides and for the SNSPDs. 

The most commonly used superconductors for SNSPDs are niobium nitride (NbN) and niobium titanium nitride (NbTiN) \cite{Goltsman2009-ur,natarajan_tanner_hadfield_2012,esmaeil_zadeh_chang_los_gyger_elshaari_steinhauer_dorenbos_zwiller_2021}. This material provides outstanding metrics, such as system detection-efficiencies above 90\% \cite{Zhang2017}, timing jitters below 5~ps \cite{korzh_zhao_allmaras_frasca_autry_bersin_beyer_briggs_bumble_colangelo_et_al._2020}, repetition rates of 25~MHz, \cite{Esmaeil_Zadeh2017-rv} and dark count-rates below 10~Hz \cite{Zhang2017,ferrari_schuck_pernice_2018}. Furthermore, the critical temperature is about 10~K even in thin-film form such that NbN-based SNSPDs operate with low noise in helium-4 cryostats at about 2~K. The integration of NbN SNSPDs with an optical waveguide results in a coupling efficiency close to unity \cite{pernice_schuck_minaeva_li_goltsman_sergienko_tang_2012}. In the past, waveguide-integrated NbN SNSPDs have been realized for several photonic platforms \cite{ferrari_schuck_pernice_2018} including GaAs \cite{reithmaier_kaniber_flassig_lichtmannecker_muller_andrejew_vuckovic_gross_finley_2015, reithmaier_lichtmannecker_reichert_hasch_muller_bichler_gross_finley_2013}. However, NbN is a single crystal. Inhomogeneities in the crystal phase of the films limit the detector yield, particularly for large-area structures, even on sapphire that is closely latticed matched \cite{gaudio_opt_hoog_zhou_sahin_fiore_2014}. Generally, any mismatch between the lattice constants limits the compatibility with potential substrates \cite{ferrari_schuck_pernice_2018,Rhazi2021dz,Tanner2012ws}. For most substrates, technologically challenging high-temperature deposition is necessary to reduce structural defects \cite{guziewicz_slysz_borysiewicz_kruszka_sidor_juchniewicz_golaszewska_domagala_al._2011, tanner_alvarez_jiang_warburton_barber_hadfield_2012}. High-yield integration of NbN SNSPDs with GaAs is therefore problematic.

Amorphous superconductors such as tungsten silicide (WSi) or molybdenum silicide (MoSi) represent an alternative material system for SNSPDs \cite{marsili_verma_stern_harrington_lita_gerrits_vayshenker_baek_shaw_mirin_et_al._2013,Verma2015jx}. These materials provide a yield close to unity on silicon substrates, even for deposition at room temperature \cite{esmaeil_zadeh_chang_los_gyger_elshaari_steinhauer_dorenbos_zwiller_2021, ferrari_schuck_pernice_2018}. The detector efficiency can be very high, above 90\%. The critical temperature is about 5~K in thin-film form and because of this, operation at a temperature of 1 K or less typically provides better performance \cite{Verma2015jx}. The amorphous nature of the materials renders them much less substrate-specific than the crystalline NbN. Recently, MoSi SNSPDs were integrated with nanophotonic silicon
nitride $\left(\text{Si}_3\text{N}_4\right)$ waveguides on a silicon chip and an on-chip efficiency of 73\% was achieved \cite{haussler_mikhailov_wolff_schuck_2020}.

An open question is the extent to which MoSi is compatible with a GaAs substrate. We answer this question here by establishing a fabrication process of MoSi-based SNSPDs on GaAs. We present a detailed analysis of the properties of the material itself and an optimization of its deposition conditions. MoSi films on a GaAs substrate superconduct with a critical temperature only slightly lower than equivalent films on a Si substrate. We successfully operate MoSi meanderlines on GaAs as single-photon detectors. The efficiency, timing jitter and recovery time of these MoSi SNSPDs on GaAs are very similar to commercial MoSi SNSPDs fabricated on silicon substrates.

\section{Optimization of superconducting {MoSi} thin-films}\label{Section2}
\begin{figure}[t!]
\includegraphics{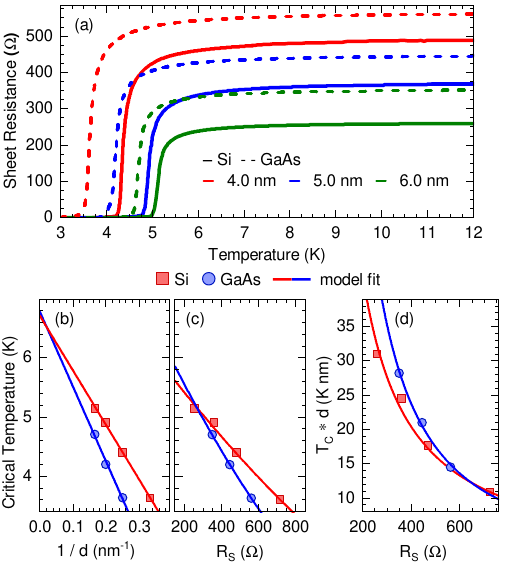}
\caption{Superconducting transition temperature and sheet resistance in the resistive phase versus film thickness for films on Si and GaAs substrates. (a) Sheet resistance versus temperature for $\text{Mo}_{0.69}\text{Si}_{0.31}$ layers of three different thicknesses deposited on a silicon surface (solid) and a natural GaAs surface (dashed). (b) Critical temperature $T_C$ versus reciprocal of the MoSi-layer thickness $d$ for deposition on Si (squares, red) and GaAs substrates (circles, blue). Data points are extracted from (a) and fitted using a function of the form $T_c = T_{co}(1-d_c/d)$ according to the Simonin's model \cite{simonin_1986}, (Eq.~\ref{jsimonin}). On Si: $T_{co}$ = 6.71 $\pm$ 0.07~K, $d_c$ = 1.38 $\pm$ 0.03~nm; on GaAs: $T_{co} = 6.79 \pm 0.23~$K, $d_c = 1.87 \pm 0.10~$nm. (c) Critical temperature $T_C$ versus sheet resistance $R_S$ extracted from (a) and fitted according to the Finkel'stein's model \cite{finkelstein_1994} (Eq.~\ref{Equation_Finkelstein}). On Si: $T_{co} = 6.22 \pm 0.07$~K, $\gamma = 7.40 \pm 0.06$; on GaAs: $T_{co} = 6.82 \pm 0.03$~K, $\gamma = 8,35 \pm 0.02$. (d) $T_C\times d$ versus $R_S$ for different thicknesses of MoSi deposited on Si (red, square) and on GaAs (blue, circle) extracted from (a). The data are fitted to Ivry's universal scaling law (Eq.~\ref{scaling_law}).}
\label{figure_1}
\end{figure}
One of the critical steps in SNSPD fabrication is the thin-film deposition of the superconducting material. MoSi is deposited via co-sputtering using a DC and RF bias for the molybdenum and silicon targets, respectively, under ultrahigh-vacuum conditions (pressure below $10^{-9}$~mbar). The target is at room temperature. Before opening the sample shutter, pre-sputtering for 60~s cleans the targets and stabilizes the sputtering conditions. Once the thickness of the MoSi film reaches the desired thickness, we close the Mo-shutter separately and continue the deposition of silicon to create a Si capping layer. Controlling the sputtering rates of Mo and Si separately allows thin films with different compositions to be deposited deterministically.

To assess the properties of the thin films, the superconducting material is cooled down to cryogenic temperatures and its resistivity is measured. A variable temperature insert (VTI) allows control of the temperature down to 1.4~K. To measure the sheet resistance of the thin layers, we use a four-point method which is independent of the contact resistance \cite{yoshimoto_murata_kubo_tomita_motoyoshi_kimura_okino_hobara_matsuda_honda_et_al._2007}.

Given a specific substrate material and sputtering process, the superconducting properties of the thin films depend on thickness and composition. The impact of the former on the superconducting transition is illustrated in Figure \ref{figure_1} for two different substrate materials, commercial intrinsic (001) silicon with a 285~nm thick oxide layer, and GaAs, specifically a 50-nm-thick GaAs epilayer grown by molecular-beam epitaxy on top of an AlAs/GaAs superlattice itself on a commercial (001) GaAs substrate. Figure \ref{figure_1}a depicts sheet resistance versus temperature curves for film thicknesses of 4.0~nm (red), 5.0~nm (blue) and 6.0~nm (green) deposited on Si (solid) and GaAs substrates (dashed). For Si, we obtain critical temperatures of 4.4~K, 4.9~K and 5.1~K; for GaAs 3.6~K, 4.2~K and 4.7~K, respectively. The critical temperatures on GaAs are slightly lower compared to those on Si. This is a consequence of the complex native oxide that forms on the GaAs surface: it  consists of a mixture of Ga- and As-related oxide compounds \cite{asaoka_2003}. This point is proven by passivating the GaAs substrate: after removal of the natural oxide, a 10-nm-thin alumina ($\text{Al}_{2}\text{O}_{3}$) layer is deposited by atomic-layer deposition \cite{najer_tomm_javadi_korsch_petrak_riedel_dolique_valentin_schott_wieck_et_al._2021}. MoSi deposited on a passivated GaAs substrate exhibits an identical critical temperature to that of MoSi on a Si substrate.

Based on Ginzburg-Landau theory, J.~Simonin developed a model describing the correlation between critical temperature $T_c$ and film thickness $d$ of a superconducting material \cite{simonin_1986}:
\begin{equation}
\label{jsimonin}
T_c = T_{co}(1-d_c/d),
\end{equation}
where $T_{co}$ is the critical temperature of the bulk material, and $d_c$ the critical thickness below which the superconducting properties of the material disappear. The critical temperatures for different thicknesses can be extracted from Figure \ref{figure_1}a and are shown in Figure~\ref{figure_1}b, a plot of $T_c$ versus $1/d$. Fitting the Simonin's model to the data results in \mbox{$T_{co}=6.71 \pm 0.07$~K} and \mbox{$d_c = 1.38 \pm 0.03$~nm} for deposition on the Si substrate; \mbox{$T_{co} = 6.79 \pm 0.23$~K} and \mbox{$d_c = 1.87 \pm 0.10$~nm} for the GaAs substrate. The intercept with the y-axis represents the critical temperature of bulk material and is identical (within the uncertainties in the fit parameters) for Si and GaAs. This can be explained by the fact that the influence of the substrate decreases with increasing thickness. Starting from bulk material, the critical temperature decreases for decreasing thickness, more rapidly for GaAs than for Si.

A.~Finkel'stein proposed a law connecting the critical temperature $T_c$ with the sheet resistance $R_s$ in the resistive phase \cite{finkelstein_1994}:
\begin{equation}
\label{Equation_Finkelstein}
\frac{T_c}{T_{co}}= \exp\left(\gamma\right)\left(\frac{1-\chi}{1+\chi}\right)^{1/\sqrt{2r}},
\end{equation}
where
\begin{equation}
\chi=\left(\frac{\sqrt{r/2}}{r/4+1/\gamma}\right)\ \text{and}\ r= \frac{e^2}{2\pi^2\hbar}R_s.
\end{equation}
$\hbar$ is the reduced Planck constant, $e$ the elementary charge, and $\gamma$ a fit parameter. This model was originally developed for quasi-two-dimensional films. However, it has been shown that the model gives realistic values for the mean scattering time, the diffusion constant, and the electron density even for films with a thickness larger than the mean free path \cite{banerjee_baker_doye_nord_heath_erotokritou_bosworth_barber_maclaren_hadfield_et_al._2017}. Figure \ref{figure_1}c shows fits of the data to Finkel'stein's law. We obtain values of $T_{co} = 6.22\pm 0.07$~K, $\gamma = 7.40 \pm 0.06$ for MoSi on a Si substrate; $T_{co}=6.82 \pm 0.03$~K and $\gamma = 8.35 \pm 0.02$ on a GaAs substrate. These high values of $\gamma$ indicate an amorphous structure of the superconducting layer \cite{finkelstein_1994}. A.~Banerjee \textit{et al.} extracted values of $T_{co} = 7.8$~K and $\gamma = 7.66 \pm 0.1$ for $\text{Mo}_{0.83}\text{Si}_{0.17}$ \cite{banerjee_baker_doye_nord_heath_erotokritou_bosworth_barber_maclaren_hadfield_et_al._2017}. The higher value of $T_c$ with respect to our results is a consequence of the higher concentration of Mo in the material.

For thin-films near the superconducting-to-resistive transition, the connection between the critical temperature $T_c$, the film thickness $d$, and the sheet resistance $R_s$ can be described by a universal scaling law \cite{ivry_kim_dane_de_fazio_mccaughan_sunter_zhao_berggren_2014}:
\begin{equation}
\label{scaling_law}
T_cd=AR_s^{-B},
\end{equation}
where $A$ and $B$ are fit parameters. Figure \ref{figure_1}d shows how the data fit to the universal scaling law. We obtain $A=11000 \pm 7000$, $B=1.06 \pm 0.11$ for MoSi deposition on a Si substrate; $A=90000 \pm 50000$ and $B=1.37 \pm 0.10$ on a GaAs substrate. ($A$ and $B$ are given in S.I.\ units.) Values of $B>1$ indicate an amorphous structure \cite{ivry_kim_dane_de_fazio_mccaughan_sunter_zhao_berggren_2014}. Small uncertainties in the thicknesses of the MoSi films have a huge influence on $T_cd$ such that $A$ cannot be determined more precisely.

\begin{figure}[t!]
\includegraphics{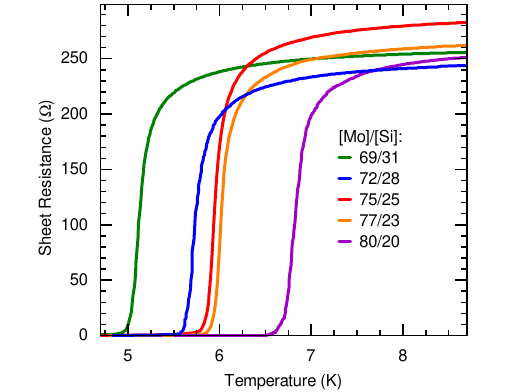}
\caption{Sheet resistance versus temperature of MoSi films with five different compositions. The films have thickness 6.0~nm and were deposited on Si substrates.}
\label{figure_15}
\end{figure}

The second parameter that affects the superconducting transition is the composition of the thin films. The composition can be controlled by setting the sputtering parameters via:
\begin{equation}
\frac{n_\text{Mo}}{n_\text{Si}} = \frac{\dot{d}_\text{Mo}}{\dot{d}_\text{Si}}\underbrace{\frac{\rho_\text{Mo}}{\rho_\text{Si}}\cdot\frac{M_\text{Si}}{M_\text{Mo}}}_{=1.286},
\end{equation}
where $\dot{d}_\text{Mo/Si}$ are the deposition rates, $\rho_\text{Mo/Si}$ the densities, and $M_\text{Mo/Si}$ the molar masses of Mo/Si. Figure \ref{figure_15} depicts sheet resistance versus temperature curves for five different compositions ranging from $\text{Mo}_{0.69}\text{Si}_{0.31}$ to $\text{Mo}_{0.80}\text{Si}_{0.20}$. The films have thickness 6.0~nm and were deposited on Si substrates. The curve representing the composition $\text{Mo}_{0.69}\text{Si}_{0.31}$ is identical to the one shown in Figure~\ref{figure_1}a (green curve). The critical temperature increases with Mo concentration from 5.1~K to 6.9~K, whereas the sheet resistance in the resistive state remains approximately constant. Interestingly, there is a range around $\text{Mo}_{0.75}\text{Si}_{0.25}$ where the critical temperature is hardly dependent on the composition. This means that the superconducting properties are weakly dependent on fluctuations in the sputtering process, which is advantageous for repeatability, also production yield. However, we decided to use a lower fraction of Mo (69\%) for the detectors to decrease the superconducting energy gap, resulting in a high internal quantum efficiency, in particular at elevated operation temperature \cite{haussler_mikhailov_wolff_schuck_2020}. For Mo concentration exceeding 80\%, a regime is approached where spontaneous crystallization can take place above a certain layer thickness, causing a sharp drop of the critical temperature \cite{adriana_e_lita_2021,Bosworth2015md}.

Similar studies can be found in the literature \cite{Zhang2021wb,Bosworth2015md,banerjee_baker_doye_nord_heath_erotokritou_bosworth_barber_maclaren_hadfield_et_al._2017,Bao2021um}. However, our results are not directly comparable with the observations in these publications since the exact details of the superconducting films differ in thickness and/or composition.

All these results demonstrate that thin films of MoSi on a GaAs substrate become superconducting at low temperature with a critical temperature only slightly impaired with respect to equivalent films on a Si substrate.

\section{Fabrication of meander-type SNSPDs on gallium arsenide substrates}
\begin{figure}[t!]
\includegraphics{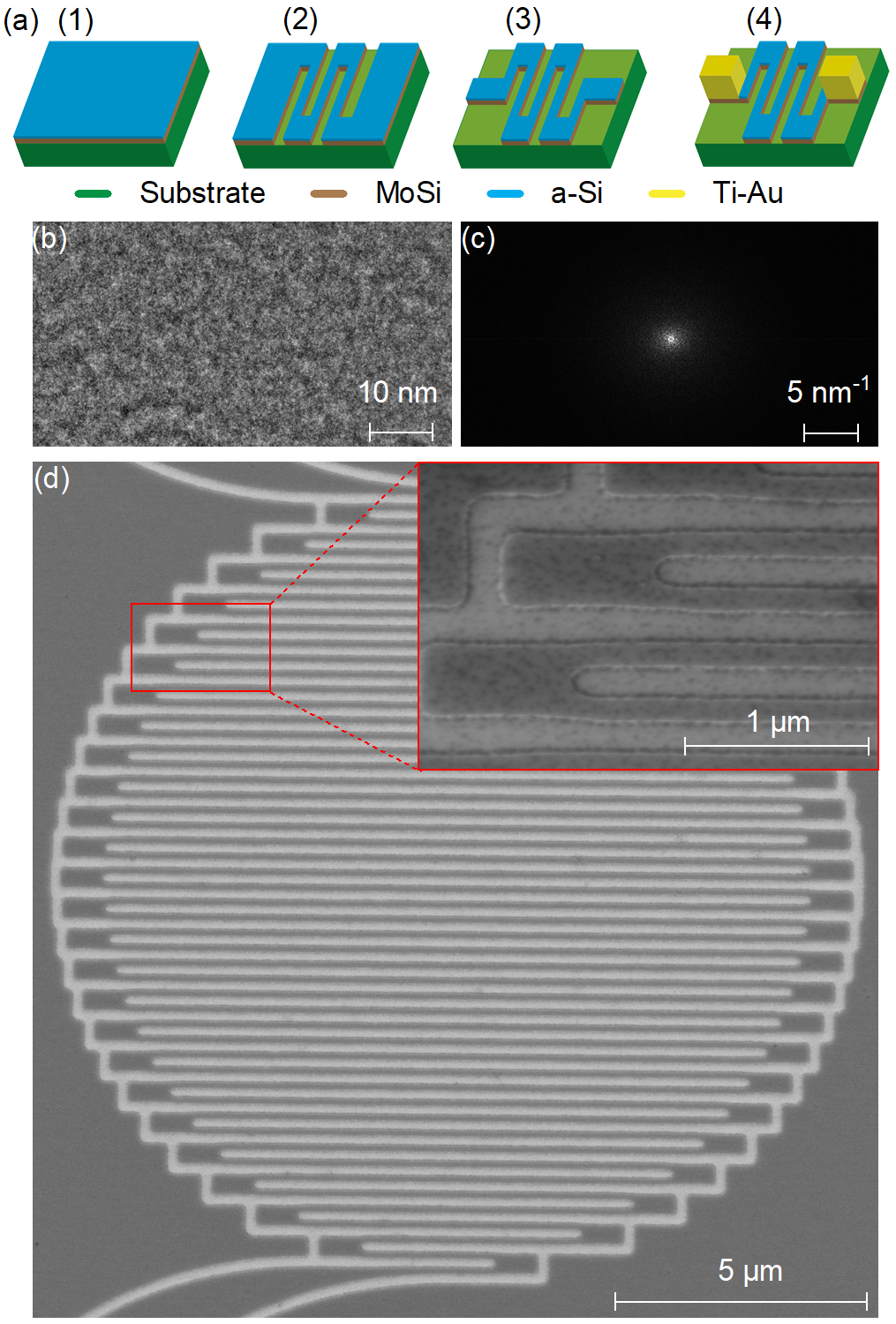}
\caption{(a) Process flow for SNSPD fabrication. (1) Deposition of the superconductor. (2) Nanowire patterning. (3) Device isolation. (4) Partial removal of protective Si layer and deposition of contact pads. (b) TEM image of a 6-nm-thick layer of MoSi capped by a 3-nm-thick layer of amorphous silicon. (c) Fourier transform of b. (d) SEM image of a nanowire featuring a width of 130~nm. Dark grey: superconductor, light grey: substrate.}
\label{figure_2}
\end{figure}
We fabricate SNSPDs on GaAs. The nanowire has a meander design, and photons are coupled from the top via a single-mode optical fibre. Figure \ref{figure_2}a sketches the main steps in the SNSPD fabrication. The starting point for the fabrication process is a GaAs wafer (green).

The first step (1) is the deposition of a 6-nm-thick $\text{Mo}_{0.69}\text{Si}_{0.31}$ film (brown) followed by a 3-nm-thick amorphous silicon capping layer (blue) that prevents the superconductor from oxidation. To prove the amorphous character of the film, we investigate the MoSi layer in a transmission-electron microscope (TEM). For this purpose, we deposit the MoSi film on pioloform coated 400 mesh copper grids that are transparent for electron beams. Figure \ref{figure_2}b shows a TEM image recorded with an acceleration voltage of 200~kV and a magnification of 500~k. The corresponding Fourier transformation is depicted in Figure~\ref{figure_2}c and shows clearly the amorphous character of the superconducting film. To verify the composition, we performed an X-ray photoelectron spectroscopy measurement.

The second step (2) is to pattern the MoSi layer into meander nanowires covering an area of 8~\textmu m radius. This is carried out via electron-beam lithography and inductively coupled plasma (ICP) etching. The nanowires are etched using an SF6/Ar plasma. The addition of Ar ions leads to physical etching, also sputtering, which reduces the edge roughness of the etched structures \cite{midolo_pregnolato_kirsanske_stobbe_2015}. Each chip contains nanowires featuring different widths (100~nm -- 150~nm) and fill factors. Figure \ref{figure_2}d shows a scanning-electron microscope image of a nanowire featuring a width of 130~nm and a gap of 120~nm. At this point, the superconducting material covers also the space in-between the individual nanowires such that all the detectors are short-circuited. To overcome this issue, the superconducting material is removed between the contact pads of the different SNSPDs using UV lithography . We call this step device isolation, step (3). The final step (4) is the fabrication of the electrode pads consisting of titanium and gold (orange) with thicknesses of 10~nm and 90~nm, respectively. To provide an electrical contact with low resistance, the amorphous silicon protection layer is locally removed beforehand via Ar milling.

\begin{figure}[t!]
\includegraphics[width=8.5cm]{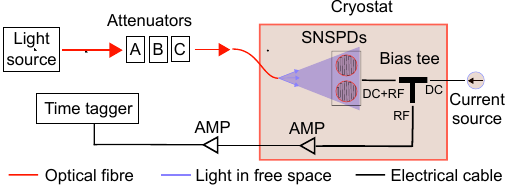}
\caption{Overview of the flood illumination setup used to test the single-photon detectors. The output signal of a cw laser is attenuated to the single-photon level by 3 variable attenuators A, B and C. The detectors are placed in a 3-stage cryostat at temperatures between 0.9~K and 2.2~K. The electrical output signal of the detection circuit is fed to a time-tagging machine. AMP: amplifier.}
\label{figure_25}
\end{figure}

\begin{figure*}[t!]
\includegraphics[width=16cm]{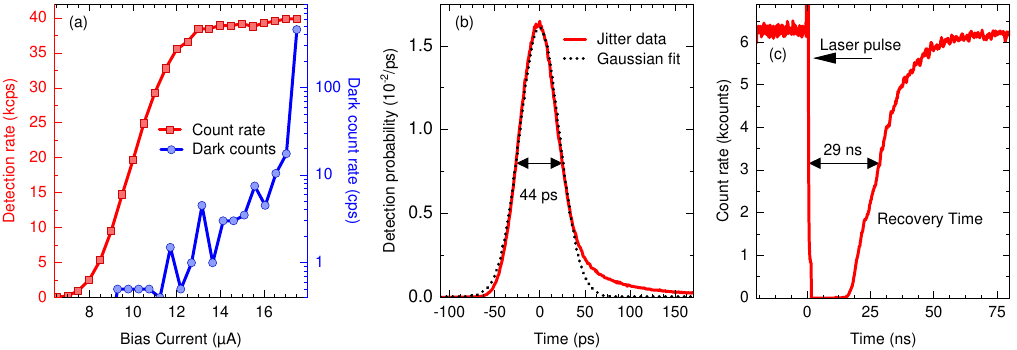}
\caption{Single-photon detector test on passivated GaAs at a wavelength of 980~nm. (a) Detection rate (red, squares) and dark count-rate (blue, circles) as a function of the bias current. (b) Jitter and (c) recovery time measurements.}
\label{figure_3}
\end{figure*}

\section{Testing of detector efficiency, timing jitter and recovery time}
\label{Testing of detector efficiency, timing jitter and recovery time}

We fabricated SNSPDs on two different types of substrates, a GaAs substrate with a native oxide at the surface and a passivated GaAs substrate (see Section \ref{Section2}). The superconducting material is a 6-nm-thick layer of MoSi (composition: 69/31) protected by a 3-nm-thick amorphous silicon layer. The SNSPDs are characterized with a flood-illuminated setup (see Figure \ref{figure_25}). The SNSPD is placed in a 3-stage cryostat and is cooled down to 0.9~K. The electrical readout circuit contains a bias tee, allowing us to apply the direct current of a constant current source while measuring the RF output signal of the detector. The voltage signal from the AC port is amplified twice, at cryogenic temperature and subsequently at room temperature, and is fed to a time-tagging machine.

\begin{table*}
\caption{Summary of the different detector tests}
\begin{tabular}{ccccccc}\hline\hline
wavelength (nm) & GaAs substrate & temperature (K) & bias current (\textmu A) & efficiency & jitter (ps) & recovery time (ns)\\ \hline
980 & native & 0.9 & 14.0 & saturated & 60 & 56 \\
980 & passivated & 0.9 & 16.5 & saturated & 50 & 29\\
1550 & native & 0.9 & 14.0 & not saturated & 64 & 73\\
1550 & passivated & 0.9 & 16.5 & almost saturated & 55 & 65\\ \hline
980 & passivated & 0.9 & 16.5 & saturated & 50 & 28\\
980/1064 & passivated & 1.5 & 15 & saturated & 56 & 34\\
980/1064 & passivated & 2.2 & 11.5 & saturated & - & 49\\ \hline\hline
\end{tabular}
\label{table_tests}
\end{table*}

The device functions as a single-photon detector -- a series of voltage pulses is created at the output in the presence of the weak optical excitation. Without the optical excitation, the count rate, i.e., the dark count-rate, is very small. To measure the detection rate, we employ a continuous wave (cw) laser which is attenuated by three variable optical attenuators (A, B, C) so that the detector is illuminated by single photons. The signal-versus-current curve is shown in Figure \ref{figure_3}a. It exhibits a plateau at large currents (red curve, wavelength $\lambda = 980$~nm, temperature $T = 0.9$~K, passivated GaAs). The plateau is important -- it demonstrates that the internal quantum efficiency is only limited by properties of the material and not by the temperature or fabrication errors \cite{Zhang2017}. The plateau can therefore be used as a proxy to determine the quantum efficiency of the MoSi-on-GaAs detector.

It is known that nanowires of this MoSi material on silicon give high detection efficiencies: absorption of a photon results in a resistive region and hence a voltage pulse at the output with almost certainty provided the material is biased in the plateau region \cite{caloz_perrenoud_autebert_korzh_weiss_schonenberger_warburton_zbinden_bussieres_2018}. We argue therefore that on GaAs, it is possible to achieve efficiencies as good at the ones on silicon, above 80\%, if the typical SNDPD structure is made, including a mirror and a cavity \cite{marsili_verma_stern_harrington_lita_gerrits_vayshenker_baek_shaw_mirin_et_al._2013,caloz_perrenoud_autebert_korzh_weiss_schonenberger_warburton_zbinden_bussieres_2018}. 

The dark count-rate increases drastically for currents above 16~µA. The origin of this behaviour are sections along the nanowire with a reduced cross section, named constrictions \cite{gaudio_opt_hoog_zhou_sahin_fiore_2014, natarajan_tanner_hadfield_2012}. Blue circles in Figure~\ref{figure_3}a indicate the dark count-rate plotted on a logarithmic scale. The critical current in this case is 18~µA.

We assess the timing uncertainty, the timing jitter, of our fibre-coupled detectors. For this, we plug a pulsed picosecond-laser (wavelength 1064~nm) and record the count rate in histogram mode as shown in Figure \ref{figure_3}b. The jitter is defined as the full-width-at-half-maximum (FWHM) and is determined to be 50~ps for a detector on the passivated GaAs substrate. The measurement shows a non-gaussian tail which becomes more pronounced at higher detection wavelengths. This behaviour was already reported \cite{caloz_korzh_ramirez_schonenberger_warburton_zbinden_shaw_bussieres_2019, caloz_perrenoud_autebert_korzh_weiss_schonenberger_warburton_zbinden_bussieres_2018}, but its origin remains unknown.

The recovery time is defined as the time required following a detection event for the detector to recover half of its efficiency \cite{Bienfang2023hq}. It can be determined using a hybrid-autocorrelation method \cite{autebert_gras_amri_perrenoud_caloz_zbinden_bussieres_2020}. This requires both a pulsed laser and a cw laser. The pulses contain a few tens of photons per pulse such that each pulse results in a detector click with a high probability. First, the pulsed laser triggers an initial detection event, after which the cw laser makes the detector click at a random time. We record the count rate in dependence on the time relative to the initial detection event at time $t = 0$. The result of this measurement (passivated GaAs substrate, wavelength 980~nm) is shown in Figure~\ref{figure_3}c. The recovery time is determined to be 29~ns.

Table \ref{table_tests} (upper half) shows an overview of the results of the detector tests differing in laser wavelength and substrate material. All the tested detectors feature an identical nanowire design. On passivated GaAs, the plateau extends from 13~\textmu A to 18~\textmu A (wavelength 980~nm). At a detection wavelength of 1550~nm, the count rate starts to rise at higher bias currents compared to the response shown in Figure~\ref{figure_3}a. Even at 1550~nm wavelength, the onset of the plateau is visible. Detection using MoSi on native oxide GaAs works slightly worse in the sense that the relative length of the plateau is shorter (19\% with respect to the bias current $I_C$ where $I_C$ is defined as the current at which the dark count-rate reaches 1~kcps) compared to the detector on the passivated substrate (relative plateau length: 26\%). As a consequence, a plateau is not observed at 1550~nm wavelength using native GaAs. For the detection at wavelength 980~nm, we obtain a timing jitter of 64~ps on native GaAs, 50~ps on passivated GaAs. The corresponding recovery times are 56~ns and 29~ns, respectively. This significant difference is due to the shape of the detection rate curve with respect to the bias current: this shows a longer plateau in the case of the passivated sample. Hence the bias current can reflow more quickly after a detection event to a value where the efficiency is maximal \cite{autebert_gras_amri_perrenoud_caloz_zbinden_bussieres_2020}. The test at wavelength 1550~nm results in timing jitters of 65~ps and 55~ps, and recovery times of 73~ns and 65~ns, on native GaAs and passivated GaAs, respectively.

These results show that passivating the GaAs surface with aluminium oxide ($\text{Al}_{2}\text{O}_{3}$) improves the detector performance. However, also the detectors on native GaAs have excellent properties, especially for wavelength 980~nm and below. These wavelengths are important for the on-chip detection of single photons emitted by InAs quantum dots.

InAs quantum dots in GaAs can operate at elevated temperatures of 4.2~K. Therefore it's also interesting to see how the detectors behave at higher temperature. Table~\ref{table_tests} (lower half) shows a detector test at temperatures of 0.9~K, 1.5~K and 2.2~K, performed on an SNSPD on passivated GaAs. The incident photons have wavelength of 980~nm for the measurements of efficiency and recovery time, 1064~nm for the jitter measurement. The data for detection at 0.9~K are identical to those in Figure \ref{figure_3}a and show a saturated internal quantum efficiency for currents between 14~\textmu A and 17~\textmu A. Detection at a higher temperature of 1.5~K results in a shorter plateau, extending from 14~\textmu A to 15.5~\textmu A. At the even higher temperature of 2.2~K, only the onset of the plateau is visible. Additionally, the counts-versus-current curve is shifted towards lower bias currents. The reason for this is that the phase transition depends on both temperature and electrical current: the critical current decreases with increasing temperature. For the timing jitter, we determine a value of 50~ps at a temperature of 0.9~K and 56~ps at 1.5~K. The recovery time is 28~ns at a temperature of 0.9~K, 34~ns and 49~ns at 1.5~K and 2.2~K, respectively. Overall, we note that the SNSPD performance does not decrease significantly from 0.9~K to 1.5~K and we achieve saturated internal quantum efficiency even at 2.2~K. This obviates the necessity of a sub-Kelvin system, at least at the near infrared wavelengths relevant for quantum dot photons.

\section{Conclusions and outlook}
We have carried out ultrahigh-vacuum sputtering of thin-film MoSi on both silicon and GaAs substrates. We find that:

(i) Thin MoSi films deposited on the native GaAs oxide superconduct but with a slightly lower critical temperature than MoSi on silicon.

(ii) MoSi on silicon has a composition-independent critical temperature for Mo concentrations between 72\% and 77\%, for film thicknesses of 6~nm.

(iii) MoSi deposited on a passivated GaAs surface (passivation layer $\text{Al}_{2}\text{O}_{3}$) has the same critical temperature as MoSi deposited on silicon.

For all three substrates (silicon, passivated GaAs, native GaAs), the superconducting properties follow the standard scaling with respect to film thickness and resistivity above the critical temperature.

The MoSi thin films were used in the fabrication of SNSPDs in a meander-wire geometry. We find that:

(a) MoSi SNSPDs on both passivated and native GaAs demonstrate a plateau in the signal--bias response at wavelength 980~nm indicating an excellent quantum efficiency. A plateau is observed up to a temperature of 2.2~K.

(b) MoSi SNSPDs on passivated GaAs have a higher critical current and a higher plateau extent than those on native GaAs.

(c) The SNSPD metrics (quantum efficiency, dark count-rate, timing jitter, recovery time) of MoSi SNSPDs on GaAs are state-of-the-art and match those of co-fabricated MoSi SNSPDs on Si.

We thereby demonstrate that MoSi, an amorphous superconductor, works well as a SNSPD on GaAs. The next step is to integrate SNSPDs with GaAs photonic circuits.

\section{Acknowledgments}
We thank Markus Wyss (Nano Imaging Lab of the Swiss Nanoscience Institute) for help with the SEM and TEM analysis and Laurent Marot (Department of Physics, University of Basel) for help with the XPS. This work received funding from the European Union’s Horizon 2020 Research and Innovation Programme under the Marie Skłodowska-Curie grant agreement No.\ 861097 (QUDOT-TECH). MS received funding from the Mercur Foundation (grant Pe-2019-0022).




\nocite{*}

\end{document}